\documentclass[12pt]{article}
   \usepackage{amssymb}
   \usepackage{amsmath}
   \usepackage{subfigure}
         \usepackage{epsfig}
         \def\href#1#2{#2}
         \def\IP{\relax{\rm I\kern-.18em P}}
         \newcommand{\beq}{\begin{equation}} 
         \newcommand{\eeq}{\end{equation}}
         \newcommand{\beqa}{\begin{eqnarray}}
         \newcommand{\eeqa}{\end{eqnarray}}

         \def\be{ \begin{equation}}\def\ee{ \end{equation}}
         \def\ba{ \begin{eqnarray}}\def\ea{ \end{eqnarray}}

         \def\cedille#1{\setbox0=\hbox{#1}\ifdim\ht0=1ex \accent'30 #1%
          \else{\ooalign{\hidewidth\char'30\hidewidth\crcr\unbox0}}\fi}
         
\begin{document}

{}~ \hfill\vbox{\hbox{hep-th/0405102}
\hbox{UUITP-14/04}
}\break

\vskip 1.cm

\centerline{\large \bf Pulsating Strings on $AdS_5 \times S^5$}
\vspace*{1.5ex}

\vspace*{4.0ex}

\centerline{\large \rm M. Smedb\"ack\footnote{
mikael.smedback@teorfys.uu.se}}
\vspace*{2.5ex}
\centerline{\large \it Department of Theoretical Physics}
\centerline{\large \it Box 803, SE-751 08 Uppsala, Sweden}
\vspace*{3.0ex}

\centerline{\large July 12, 2004}
\vspace*{3.0ex}

\vspace*{4.5ex}
\medskip
\bigskip\bigskip
\centerline {\bf Abstract}

We find the anomalous dimension and the conserved charges of an R-charged 
string 
pulsating on $AdS_5$.
The analysis is performed both on the gauge and string side,
where we find agreement at the one-loop level.
Furthermore, the solution is shown to be related by analytic continuation 
to a string which is pulsating on $S^5$, thus providing an example of
the close relationship between the respective isometry groups.
         
\bigskip

\vfill \eject
\baselineskip=17pt



         \section{Introduction}\label{intro}
The AdS/CFT conjecture \cite{AdS1,AdS2,AdS3}
has lead to a better understanding of both conformal gauge theories 
as well as string theory in curved spaces.
Within this framework, the seminal work of \cite{BMN} included a discussion of
operators of the form Tr$Z^{J_1}W^{J_2}+\cdots$ (built up from the scalars
$Z$ and $W$ of the $\mathcal{N}=4$ SYM supermultiplet) 
where $J_1 \ll J_2$. 
The dots indicate other permutations of the fields $Z$ and $W$ inside
the trace, and in general these states mix among themselves under scaling; 
only certain linear combinations are eigenstates to the scaling
operator.
Semiclassical string configurations which usually go beyond
the BMN limit (e.g. by taking both $J_1$ and $J_2$ to be large)
have since been studied extensively
\cite{semiclass1,semiclass2,semiclass3,semiclass4,semiclass5,
semiclass6,semiclass7,semiclass8,semiclass9,semiclass10,
semiclass11,semiclass12,semiclass14,semiclass15}
(see also \cite{scalso1,scalso2,scalso3,scalso4}),
and are 
reviewed in
\cite{semiclass13}.

The observation of \cite{YM1} that the matrix of anomalous 
dimensions
could be mapped to an integrable Bethe spin chain \cite{bethe}
simplified and extended the studies of the corresponding gauge theory  
\cite{YM2,YM3,YM4,YM7,YM8,YM9,YM10,YM12,YM13,YM15,YM17,YM18}
(see also \cite{YMalso1,YMalso2}),
reviewed in \cite{faddeev}. The original results of \cite{YM1} were
restricted to the group $SO(6)$ at 1-loop level, but were later
extended to the full 1-loop $SU(2,2|4)$ chain \cite{YM5,YM6},
taking advantage of previous results on integrability in QCD
amplitudes \cite{QCD1,QCD2} and the QCD dilatation operator
\cite{QCDdilop1,QCDdilop2,QCDdilop3}
(see also \cite{QCDmixed1,QCDmixed2,QCDmixed3,
QCDintegrab1,QCDintegrab2}).
Progress on higher orders in closed subsectors has also been made 
\cite{YM2extra,YM11,YM14,YM16,priv}.

The integrable spin chain formulation exposes the conserved charges.
Conserved charges in the sigma model were first discussed in
\cite{HCC1,HCC2} (see also \cite{HCC3,HCC4}).
Progress on relating the conserved charges on either side
to each other by viewing (subsectors of) both sides of the duality as 
an integrable system was made in \cite{HCC5,HCC6,HCC7,HCC8} 
(see also \cite{Roiban}). 
The work on finding descriptions of the AdS/CFT duality
in terms of integrable systems
are reviewed in \cite{review}.

In this paper, we will analyse a string pulsating on $AdS_5$ and 
whose centre of mass is revolving on $S^5$, 
both from the gauge and string side (assuming
large quantum numbers).
From the AdS/CFT conjecture, we expect that the anomalous dimension
of the corresponding operator will coincide with the first order energy
correction on the string side.
Another motivation for studying this configuration is that 
the conserved charges on either side
of the duality can be matched explicitly using integrability.

A third motivation
is that our solution will be shown to
be related by an analytic continuation to the solution of \cite{YM9}
for a string pulsating and revolving on $S^5$. This
provides
an example of the close mathematical relationship between
the isometry groups of $AdS_5$ and $S^5$; $SO(4,2)$ and $SO(6)$,
respectively.
Such relations were discussed in \cite{YM7}, where a first example
was given; a string rotating in two planes on $S^5$ was shown to
be related by analytic continuation to a string whose centre of mass
is revolving in one plane
on $S^5$ and with one spin in $AdS_5$.
One may speculate that tying together seemingly different solutions
in this way may help in providing a bridge between
duality checks at the level of individual solutions and higher-level
checks. An example of the latter is the recent analysis of the
duality at the level of actions \cite{al1,al2,al3,al4,al5,al6}.

We will analyse the case at hand from the gauge side and string side
in sections \ref{gauge} and \ref{string}, respectively.
In section \ref{conserved} we exhibit the conserved charges on the
string side. Our conclusions are presented in section \ref{conclusions}.


\section{Gauge Side}\label{gauge}
\setcounter{equation}{0}

In this section, 
we will consider operators of the form
Tr$(D\bar{D})^B Z^J$, which are charged under $SO(2,2)$.
Here, $D \equiv D_1 +iD_2$,
(where $D_i$ are covariant derivatives) and $Z$ is one of the three
complex scalars of the $\mathcal{N}=4$ supermultiplet.
Individual operators are formed by 
linear combinations of different distributions
of the $D$'s and $\bar{D}$'s over the $Z$'s.
In general 
mixing occurs under scaling within the full $SO(4,2)$.
However, in the semiclassical limit it turns out that it
{\it will} be sufficient
to consider the bosonic subgroup $SO(2,2)$ \cite{priv},
cf. what happens in the $SO(6)$ case \cite{YM9,HCC8}.
The mapping of the matrix of anomalous
dimensions to a Hamiltonian of a spin chain will then allow us to
find the the eigenvalues of the diagonalized system.

The simple roots of $SO(2,2)$ are
$\overrightarrow{\alpha}_1 = (1,1)$ and
$\overrightarrow{\alpha}_2 = (1,-1)$.
In the infinite-dimensional representation of highest weight
$\overrightarrow{w} = (-1,0)$,
the Bethe equations are
\begin{equation}
  \left( \frac{u_{q,i}+i\overrightarrow{\alpha}_q \cdot \overrightarrow{w}/2}
              {u_{q,i}-i\overrightarrow{\alpha}_q \cdot \overrightarrow{w}/2}
  \right)^L =
  \prod_{j \neq i}^{n_q} 
    \frac{u_{q,i}-u_{q,j}+i\overrightarrow{\alpha}_q \cdot \overrightarrow{\alpha}_q/2}
         {u_{q,i}-u_{q,j}-i\overrightarrow{\alpha}_q \cdot \overrightarrow{\alpha}_q/2}
  \prod_{q' \neq q} \prod_j^{n_{q'}}
    \frac{u_{q,i}-u_{q',j}+i\overrightarrow{\alpha}_q \cdot \overrightarrow{\alpha}_q'/2}
         {u_{q,i}-u_{q',j}-i\overrightarrow{\alpha}_q \cdot \overrightarrow{\alpha}_q'/2},
\end{equation}
as written in \cite{YM1} for an arbitrary Lie group, and
\begin{equation}\label{bethe}
  \begin{split}
    \left( \frac{u_{1,i}-i/2}{u_{1,i}+i/2} \right)^L
    & = \prod_{j \neq i}^{n_1} \frac{u_{1,i}-u_{1,j}+i}{u_{1,i}-u_{1,j}-i} \\
    \left( \frac{u_{2,i}-i/2}{u_{2,i}+i/2} \right)^L
    & = \prod_{j \neq i}^{n_2} \frac{u_{2,i}-u_{2,j}+i}{u_{2,i}-u_{2,j}-i}
  \end{split}
\end{equation}
for $SO(2,2)$.
The anomalous dimension is
\begin{equation}\label{andim}
  \gamma = \frac{\lambda}{8\pi^2} \left( 
            \sum_{i=1}^{n_1} \frac{1}{u_{1,i}^2 + 1/4}
           +\sum_{i=1}^{n_2} \frac{1}{u_{2,i}^2 + 1/4} \right) .
\end{equation}
As indicated, there are $n_q$ roots of the type $q$.
The form of the operator we are looking for is 
Tr$(D\bar{D})^B Z^J$,
so the number of sites is $L=J$. 
The two root types essentially correspond to creation of 
$D$'s and $\bar{D}$'s,
respectively, so we set $n \equiv n_1 = n_2 = B$.

Assuming that the number of roots is large (so that they can
be approximated by a continuous distribution) 
in the thermodynamic 
limit (i.e. a large number of sites $L$)
the log of the Bethe equation for the first type of
root (after a rescaling $u \rightarrow uL$) is
\begin{equation}\label{thermo}
  \frac{2}{\alpha} \left( -\frac{1}{u} + 2\pi m \right) =
  2 -\hspace{-0.48cm}\int_C \frac{\sigma(u')du'}{u-u'},
\end{equation}
where the line through the integral sign 
indicates that the singularity at $u'=u$ is resolved by taking
the principal value of the integral.
The contour $C$ is defined by the support of the root density $\sigma(u')$
and its endpoints are $a$ and $b$.
We have defined $\alpha \equiv n/L$. The integer $m$
corresponds to different branches of the $\log$. 
The root density
is normalized as
\begin{equation}\label{norm}
  \int_C \sigma(u') du' = 2.
\end{equation}
Reading (\ref{thermo}) as a force balancing equation, we conclude
that the roots are repelled from each other but attracted to the point
$u=1/2 \pi m$. We therefore expect that the roots will spread out along
the contour $C$ passing through this point.

Performing an inverse Hilbert transform on (\ref{thermo}), the
root density is
\begin{equation}
  \sigma(u) = - \frac{1}{\pi^2 \alpha} \left[ (u-a)(u-b) \right]^{1/2}
    -\hspace{-0.48cm}\int_C du' \left( \frac{1}{u'} - 2\pi m \right) \frac{1}{u'-u}
    \frac{1}{ \left[(u'-a)(u'-b) \right]^{1/2} }.
\end{equation}
The multivalued function $[\cdots]^{1/2}$ has a cut along the segment
of the real axis coinciding with the contour $C$.
Calculating the integral by deforming the contour and picking
up the residue at $u'=0$, we get
\begin{equation}\label{sigma}
  \sigma(u) = - \frac{i}{\pi \alpha u \sqrt{ab}} 
                \left[ (u-a)(u-b) \right]^{1/2}.
\end{equation}

The endpoints $a$ and $b$ of the contour $C$ are determined by inserting
(\ref{sigma}) into equations (\ref{thermo}) and (\ref{norm}).
This results in the two equations
\begin{equation}
  \begin{split}
    \sqrt{ab} & = \frac{1}{2\pi m} \\
        a + b & = \frac{1+2 \alpha}{\pi m}
  \end{split}.
\end{equation}
In particular, this means that for non-negative (i.e. physical) values
of $\alpha$, the endpoints of the contour will lie on the positive
real axis (for positive $m$) and the contour will pass
through the point $u=1/2 \pi m$, as expected.

Now define the resolvent
\begin{equation}
  W(u') \equiv \int_C   du \frac{\sigma(u)}{u'-u}.
\end{equation}
By deforming the contour and picking up residues at $u=0$, $u=u'$
and $u=\infty$, the resolvent becomes \cite{YM9,HCC6}
\begin{equation}\label{resolvent}
  -\alpha W(u') = \frac{1}{u'} \left[ 1 -   
    \sqrt{(1-2\pi mu')^2-2\alpha(4\pi mu')} \right] + \pi m.
\end{equation}
The square root denotes the branch which coincides with the principal
branch for small $u'$.
One of the virtues of the resolvent is that it determines the
first part of the
anomalous dimension (\ref{andim}) in the thermodynamic limit:
\begin{equation}\label{anom}
  \gamma_1 = -\frac{\lambda \alpha}{16 \pi^2 L} W'(0)
\end{equation}
Inserting (\ref{resolvent}) into (\ref{anom}), we get
\begin{equation}
  \gamma_1 = + \frac{\lambda m^2}{2L} \alpha (1 + \alpha).
\end{equation}
According to (\ref{bethe}), the two root types behave symmetrically
and do not interact. Due to the trace condition, which in this case
takes the form
\begin{equation}\label{trace}
  \prod_{j=1}^{j=L} 
    \frac{(u_{1,j}+i/2)(u_{2,j}+i/2)}{(u_{1,j}-i/2)(u_{2,j}-i/2)} = 1,
\end{equation}
the second type of roots spread out along along
the segment $[-b,-a]$ of the negative real axis, 
so (\ref{andim}) becomes
\begin{equation}
  \gamma = \gamma_1 + \gamma_2 = 2\gamma_1 
         = + \frac{\lambda m^2}{L} \alpha (1 + \alpha)
         = + \frac{\lambda m^2 B}{J^2} \left( 1 + \frac{B}{J} \right).
\end{equation}
The conformal dimension is $\Delta = 2B+J$, in terms of which
\begin{equation}\label{anomdim}
  \gamma = m^2 \lambda \frac{\Delta^2-J^2}{4J^3}.
\end{equation}


         \section{String Side}\label{string}
         \setcounter{equation}{0}
In \cite{semiclass4}, a string pulsating on $AdS_5$ or $S^5$ was considered.
In \cite{YM9}, the latter configuration 
was generalized to include a rotation
in one plane on $S^5$.
In this section, we will consider the closely related 
sigma model description of
a string pulsating on $AdS_5$ and whose centre of mass is
revolving on $S^5$.
Restricting the motion to the subspace $AdS_3 \times S^1$
means that the isometry group contains a factor isomorphic to
$SO(2,2)$, hence matching the set of operators considered in
section \ref{gauge}.

The metric on $AdS_5 \times S^5$ will be written
\begin{equation}
  \begin{split}
  ds^2_{AdS_5} & = d\rho^2 -\cosh^2 \rho dt^2 + \sinh^2 \rho
         ( d \theta^2 + \cos^2 \theta d\Phi_1^2 + \sin^2 \theta d\Phi_2^2 ) \\
  ds^2_{S^5}   & = d \gamma^2 + \cos^2 \gamma d\phi_3^2 + \sin^2 \gamma
         ( d\Psi^2 + \cos^2 \Psi d\phi_1^2 + \sin^2 \Psi d\phi_2^2)
  \end{split}.
\end{equation}
We will use the notation $\phi \equiv \phi_3$.
The relevant metric for the $AdS_3 \times S^1$ subspace of the full space is
\begin{equation}
  ds^2 = d\phi^2 + d\rho^2 - \cosh^2 \rho dt^2 + \sinh^2 \rho d\theta^2.
\end{equation}
We assume that the string is wrapped around the azimuthal angle
$\theta$ on $AdS_3$. We then use the ansatz
\begin{equation}
  \phi=\phi(\tau), \rho=\rho(\tau), t=\tau, \theta=m \sigma.
\end{equation}
The integer $m$ allows for multi-wrapping.
We will consider $t$ and $\theta$ to be gauge fixed.
The Nambu-Goto action is
\begin{equation}
    S = -m\sqrt{\lambda} \int dt \sinh \rho 
        \sqrt{\cosh^2 \rho -\dot{\phi}^2 - \dot{\rho}^2}.
\end{equation}
The energy $\pi_t \propto H$ and the spin $\pi_\phi$ are conserved.
The dynamical momenta are
\begin{equation}
    \pi_\rho = \frac{m \sqrt{\lambda}\dot{\rho}\sinh \rho}
                     {\sqrt{\cosh^2 \rho -\dot{\phi}^2 - \dot{\rho}^2}}
\end{equation}
and
\begin{equation}
    \pi_\phi = \frac{m \sqrt{\lambda}\dot{\phi}\sinh \rho}
                     {\sqrt{\cosh^2 \rho -\dot{\phi}^2 - \dot{\rho}^2}}.
\end{equation}
The Hamiltonian is then given by
\begin{equation}
      H^2 = (\pi_\rho \dot{\rho} + \pi_{\phi} \dot{\phi} - L)^2 
          = \cosh^2 \rho (\pi_\rho^2 + \pi_\phi^2 + m^2 \lambda \sinh^2 \rho).
\end{equation}
Following \cite{YM9}, we now consider the term $V(\rho) = m^2 \lambda \cosh^2 \rho \sinh^2 \rho$ 
to be a perturbation.
A Hermitian form of the unperturbed Hamiltonian operator acting on a 
wave function is then
$\hat{H}_0^2 \Psi(\rho) = 
  \cosh \rho (\hat{\pi}_\rho^2 + \hat{\pi}_\phi^2) \cosh \rho \Psi(\rho)$, i.e.
\begin{equation}\label{startekv}
  \Delta^2 \Psi(\rho) = -(\cosh \rho) \nabla^2 (\cosh \rho) \Psi(\rho)
                      + J(J+4) \cosh^2 \rho \Psi(\rho),
\end{equation}
where
\begin{equation}
  \nabla^2 = \frac{1}{\sinh^3 \rho \cosh \rho} 
             \frac{\partial}{\partial \rho}
             \sinh^3 \rho \cosh \rho \frac{\partial}{\partial \rho}.
\end{equation}
$J$ and $\Delta$ are non-negative integers.
Introducing $x=\frac{1}{\cosh^2 \rho}$ and restricting to even
integers $J=2j$, $\Delta=2a$ transforms (\ref{startekv}) into
\begin{equation}
  -\frac{x^{7/2}}{1-x} \frac{d}{dx} \frac{(1-x)^2}{x} \frac{d}{dx}
    \frac{1}{x^{1/2}} \Psi(x) + j(j+2) \Psi(x) - a^2 \Psi(x) = 0.
\end{equation}
The power series ansatz
\begin{equation}\label{ansatz}
  \Psi(x) = \sum_{\lambda=0}^{\lambda=\infty} a_\lambda x^{k+\lambda}
\end{equation}
results in an indicial equation with two solutions. The solution
which keeps all terms finite on the interval $0 \le x \le 1$ is
$k = j+5/2 $. Hence, the recursion relation becomes
\begin{equation}
  a_\lambda = - a_{\lambda-1} \frac{(a+1+\lambda+j) (a-1-\lambda-j)}
                                   {\lambda (\lambda+2+2j)},
\end{equation}
whose solution is
\begin{equation}
  a_p = (-1)^p
  \left(\hspace{-0.20cm}
  \begin{array}{cc}
    a-j-2 \\
    p
  \end{array}
  \hspace{-0.20cm} \right)
  \frac{(a+j+1+p)!}{(p+2j+2)!}.
\end{equation}
Inserting this into (\ref{ansatz}) gives the wave functions,
whose normalized forms are
\begin{equation}
  \Psi(x) = \frac{2\sqrt{a(a-j-1)}}{(a-j-1)!\sqrt{a+j+1}}
    \frac{1}{x^{j-1/2}} \left( \frac{d}{dx} \right)^{a-j-1}x^{a+j+1}
    (1-x)^{a-j-2}.
\end{equation}
The first order correction to the energy is
\begin{equation}
  \Delta E^2 = \int d\tau \Psi(x) V(x) \Psi(x)
             = \frac{2a(a+j+1)(a-j-1)}{(j+1)(2j+1)(2j+3)},
\end{equation}
where $d\tau$ is the volume element.
The energy to first order and for large quantum numbers is then
$E=\Delta+\gamma$, where
\begin{equation}\label{energycorr}
  \gamma = m^2 \lambda \frac{\Delta^2-J^2}{4J^3}.
 \end{equation}
This agrees with the expected anomalous dimension (\ref{anomdim}).


\section{Conserved Charges}\label{conserved}

On the gauge side, the mapping of the matrix of anomalous
dimensions to a Hamiltonian of an integrable spin chain
immediately provides all conserved charges
in terms of
the resolvent. Following \cite{HCC6}, 
the recent paper \cite{HCC7} demonstrates
that for the R-charge assignment $(J_1,J_2,J_2)$ on $S^5$,
the charges on the string side precisely match those on
the gauge side (at the 1-loop level).
In our case, the 1-loop resolvent 
is (\ref{resolvent}). On the string side, the
corresponding generator is essentially given by the
quasi-momentum, as discussed in \cite{HCC8}. In this section,
we will follow the procedure outlined in \cite{HCC8} to
exhibit the quasi-momentum for the case at hand.

We are assuming that the string is moving on an $AdS_3 \times S^1$
subspace. The $AdS_3$ space can be described as the hypersurface
$-X_1^2 -X_2^2 + X_3^2 + X_4^2 = 1$ in $\mathbb{R}^4$. Defining $W \equiv X_1 + iX_2$
and $Z \equiv X_3 +iX_4$, this space can be equivalently described
as an $SU(1,1)$ group manifold using the map
\begin{equation}
  \left(
  \begin{array}{cc}
    Z       & W        \\
    \bar{W} & \bar{Z}  \\
  \end{array}
  \right)
  = g \in SU(1,1).
\end{equation}
As an ansatz for the string pulsating on $AdS_3$ and revolving on $S^1$
we use
\begin{equation}
  \begin{array}{cc}
    W & = \sinh \rho e^{i \theta} \\
    Z & = \cosh \rho e^{i t}
  \end{array}.
\end{equation}
In this section we will use the Polyakov action in unit gauge,
\begin{equation}\label{action}
  \begin{array}{cc}
    S & = \frac{\sqrt{\lambda}}{4\pi} \int d\sigma d\tau 
        \left[
          \partial Z \partial \bar{Z} - \partial W \partial \bar{W} 
          -(\partial X_5)^2
        \right] \\
      & = - \frac{\sqrt{\lambda}}{4\pi} \int d\sigma d\tau
          \left[
            \frac{1}{2} \mbox{Tr} (g^{-1}\partial_\alpha g)^2
            +(\partial X_5)^2
          \right]
  \end{array}
\end{equation}
(hence choosing $X_3$ and $X_4$ to be time-like).
In this description, we will no longer consider $t(\tau)$ to be gauge
fixed.
By the equations of motion, $\phi \equiv X_5 = Q \tau$.
The action is invariant under constant shifts along the
circle, so $Q\sqrt{\lambda} \equiv J$ is the conserved charge corresponding to the spin.

The action (\ref{action}) is also invariant under constant left and right
shifts of the group elements, $g \rightarrow hg$ and $g \rightarrow gh$.
The corresponding charge is the energy 
\begin{equation}
  E = +\sqrt{\lambda} Q_l = -\sqrt{\lambda}Q_r = -\sqrt{\lambda} \dot{t} \cosh^2 \rho.
\end{equation}
In the following, we will restrict our considerations to the
time $\tau$ when $\rho(\tau)=t(\tau)=0$. Then
\begin{equation}
    \frac{E^2}{\lambda} = \dot{t}^2 = \dot{\rho}^2 + Q^2,
\end{equation}
where the last equality follows from the constraint corresponding
to fixing the gauge in the Polyakov action.

Defining $\partial_{\pm} \equiv \partial_\tau \pm \partial_\sigma$
and currents $j_{\pm} \equiv g^{-1} \partial_{\pm} g$, it follows from the
constraint $Z\bar{Z}-W \bar{W}=\det(g)=1$ that
\begin{equation}
  0 = \partial_+ j_- - \partial_- j_+ + [j_+,j_-].
\end{equation}
This coincides with the consistency condition $[L,M]=0$ for the linear
problem $L\Psi = M\Psi = 0$, where
\begin{equation}
  \begin{array}{cc}
    L & = \partial_\sigma +\frac{1}{2} 
        \left( \frac{j_+}{1-x} - \frac{j_-}{1+x} \right) \\
    M & = \partial_\tau +\frac{1}{2} 
        \left( \frac{j_+}{1-x} + \frac{j_-}{1+x} \right)
  \end{array}.
\end{equation}
Explicitly, the first equation is
\begin{equation}\label{sigmaprob}
  \partial_\sigma \Psi = \frac{x}{x^2-1}
  \left(
  \begin{array}{cc}
    iQ_l                   & \dot{\rho}e^{im\sigma} \\
    \dot{\rho}e^{-im\sigma} & -iQ_l
  \end{array}
  \right)
  \Psi
\end{equation}
Considering $\Psi$ to be a vector of the type
\begin{equation}
  \Psi = \left(
  \begin{array}{cc}
    A e^{ip_{+} \sigma/2\pi} \\
    B e^{ip_{-} \sigma/2\pi}
  \end{array}
  \right)
\end{equation}
provides a family of solutions\footnote{
The ansatz $p_\pm(x)=a(x) \pm \pi m$ is helpful.} 
to (\ref{sigmaprob}), provided that
the condition
\begin{equation}
  \left[ \frac{x^2-1}{2\pi} (p_{\pm} \mp \pi m) \right]^2
  + (x\dot{\rho})^2
  = \left[ \frac{m(x^2-1)}{2} - xQ_l \right]^2
\end{equation}
is satisfied.
Consequently, each root $p_\pm(x)$ will be double-valued.
Subtracting the poles from one of the sheets of $p_-$
(with a branch cut along the positive real axis),
the resolvent is
\begin{equation}
  G(x) = \frac{2\pi}{x^2-1} \left[
         -xQ + \left\{ \left[ \frac{m}{2}(x^2-1)-xQ_l \right]^2 - (x\dot{\rho})^2 \right\}^{1/2}
         \right]
         -\pi m.
\end{equation}
Rescaling $x \rightarrow 4\pi Qx$, the leading contribution for large
quantum numbers ($\lambda \rightarrow 0$ in $Q=J/\sqrt{\lambda}$ and
$Q_l = E/\sqrt{\lambda}$) is
\begin{equation}\label{resolvent_string}
  -G_0(x) = \frac{1}{2x} \left[
           1-\sqrt{(1-2\pi mx)^2 - 2\alpha (4\pi mx)}
           \right] + \pi m.
\end{equation}
The square root denotes the branch which coincides with the principal
branch for small $x$.
It is proportional to the resolvent (\ref{resolvent}) on the gauge theory
side.
Since the charges are generated by the odd part of the resolvent,
this shows that the charges on the gauge and string side match.


\section{Conclusions}\label{conclusions}

We considered a string pulsating on $AdS_5$ and revolving on $S^5$.
The anomalous dimension (\ref{anomdim}) agrees with the first
order energy correction (\ref{energycorr}), as expected from
the AdS/CFT conjecture. In terms of 
$\alpha \equiv \frac{n}{J} = \frac{\Delta-J}{2J}$, these
results become
\begin{equation}\label{energycorr}
  \gamma = m^2 \lambda \frac{1}{J} \alpha (1+\alpha).
\end{equation}

Consider analytically continuing this result to the unphysical region
$\alpha < 0$ by $\Delta \rightarrow - J_1$ and
$J \rightarrow -L$. This takes 
$\alpha \rightarrow -\alpha_{EMZ} \equiv \frac{J_1-L}{2L}$, i.e.
\begin{equation}\label{energycorr}
  \gamma \rightarrow m^2 \lambda \frac{1}{L} \alpha_{EMZ} (1-\alpha_{EMZ}).
\end{equation}
This is the result\footnote{
Our definition of $\alpha_{EMZ}$ differs by a factor of 2 from that of
\cite{YM9}.}
of \cite{YM9} for a string pulsating and revolving
on $S^5$, which together with our result provides a complete
description for all real values of $\alpha$; for $\alpha < 0$,
the string pulsates and revolves on $S^5$. As $\alpha$ is turned
to positive values\footnote{
Note that we assumed large quantum numbers.
As $J \rightarrow 0$, 
the thermodynamic limit is no longer valid.
The behaviour of the anomalous dimension 
in the strong coupling region is discussed in \cite{semiclass4}.
A similar phenomenon occurs in \cite{YM7}.
}, 
the string starts pulsating on $AdS_5$ instead
(while still revolving on $S^5$).
On the gauge side, the corresponding operator forms are
Tr$(Z\bar{Z})^{(L-J_1)/2} X^{J_1}$ and
Tr$(D\bar{D})^{(\Delta-J)/2} Z^J$, respectively.

This description can be compared to the extension of \cite{YM7}
to the results of \cite{YM4}. 
In \cite{YM4}, operators of the form Tr$Z^{J_1}W^{J_2}$ were considered, 
corresponding to strings rotating in two planes on $S^5$.
It was shown in \cite{YM7} that the replacements
$J_1 + J_2 \rightarrow - J$, $J_2 \rightarrow S$ and 
$\gamma \rightarrow - \gamma$ turned the system 
of Bethe equations\footnote{
In the present case and 
on the level of Bethe equations in the thermodynamic limit, 
taking $\alpha \rightarrow -\alpha_{EMZ}$ turns (\ref{thermo})
into the corresponding equation in \cite{YM9}.
}
and anomalous dimensions
into a description
of operators of the form Tr$D^S Z^J$, corresponding to strings
rotating both on $AdS_5$ and $S^5$.

Let us also mention that the presence of an integrable structure on both 
sides of the duality
is manifested in our case by the agreement of the corresponding generators of 
conserved charges, (\ref{resolvent}) and (\ref{resolvent_string}).
For the string pulsating on $S^5$, considered in \cite{YM9}, 
the corresponding check was carried out in \cite{HCC8}.


         \bigskip
         \noindent
        {\bf Acknowledgments:}
         I am very grateful to J.\ Minahan for helpful discussions
         and support during the course of this work, and for
         useful comments
         on the manuscript. I would also like to thank  
         J.\ Engquist, L.\ Freyhult and K.\ Zarembo for conversations.



           \footnotesize

           \bibliographystyle{unsrt}
           \bibliography{references_rev2}
           



         \end{document}